\documentclass[prd,aps,twocolumn,a4paper,floatfix]{revtex4}

\usepackage{graphicx,psfrag}
\usepackage{mathrsfs}
\usepackage{amsmath,amsfonts,amssymb} 

\def\p{\partial}
\def\tnu{\tilde{\nu}}
\def\Lie{{\cal L}}
                     
\def\half{\frac{1}{2}}

\usepackage{color}
\usepackage{float}
\usepackage{ulem}
\definecolor{cyan}{rgb}{0,0.9,0.9}
\definecolor{orange}{rgb}{0.9,0.5,0}
\definecolor{magenta}{rgb}{1,0,1}
\definecolor{purple}{rgb}{0.8,0.4,0.8}

\begin{document}

\title{Constraint damping for the Z4c formulation of general 
relativity}

\author{Andreas Weyhausen, Sebastiano Bernuzzi and David Hilditch}

\address{Theoretical Physics Institute, University of 
  Jena, 07743 Jena, Germany}

\date{\today}

\begin{abstract}
One possibility for avoiding constraint violation in numerical 
relativity simulations adopting free-evolution schemes is to 
modify the continuum evolution equations so that constraint 
violations are damped away. Gundlach et. al. demonstrated that such a 
scheme damps low-amplitude, high-frequency constraint-violating modes 
exponentially for the Z4 formulation of general relativity. Here we 
analyze the effect of the damping scheme in numerical applications on 
a conformal decomposition of Z4. After reproducing the theoretically 
predicted damping rates of constraint violations in the linear regime, 
we explore numerical solutions not covered by the theoretical analysis. 
In particular we examine the effect of the damping scheme on 
low-frequency and on high-amplitude perturbations of flat spacetime 
as well and on the long-term dynamics of puncture and compact star 
initial data in the context of spherical symmetry. We find that the 
damping scheme is effective provided that the constraint violation is 
resolved on the numerical grid. On grid noise the combination of 
artificial dissipation and damping helps to suppress constraint 
violations. We find that care must be taken in choosing the damping 
parameter in simulations of puncture black holes. Otherwise the 
damping scheme can cause undesirable growth of the constraints, and 
even qualitatively incorrect evolutions. 
In the numerical evolution of a compact static star we find that the 
choice of the damping parameter is even more delicate, but may lead to a small 
decrease of constraint violation. For a large range of values it results in 
unphysical behavior.

\end{abstract}


\maketitle

\section{Introduction}
\label{sec:Introduction}

The most common way to construct numerical solutions to the 
field equations of general relativity is to take a free-evolution 
approach. The Hamiltonian and momentum constraints of the 
theory are explicitly solved only for initial data. Then the 
remaining field equations are rewritten in a suitable hyperbolic 
form, and the initial data can be evolved using the desired 
numerical method with this {\it hyperbolic formulation}. In the 
absence of boundaries the contracted Bianchi identities can be 
used to show that if the constraints are satisfied on one spacelike 
slice of a foliation, then they will be satisfied everywhere. However, 
numerical solutions violate the constraints. This violation can 
be considered under control, if when one applies more resolution 
to the problem, the constraint violation converges away at an 
appropriate rate. Nonetheless, even if the constraint violation 
converges away, at finite resolution constraint violation is 
undesirable. A number of strategies to minimize the violation have 
been considered. One is to choose the formulation such that 
every constraint propagates. In combination with suitable boundary 
conditions, the constraint violation on the numerical grid should 
then be propagated away. On the other hand, if the constraints do 
not propagate then any violation may sit on the grid and grow.
Another strategy is to use a {\it constraint damping scheme}, 
namely to modify the evolution equations so that the constraint-satisfying 
hypersurface becomes an attractor in phase space; such an evolution scheme 
is sometimes called a $\lambda$-system~\cite{Brodbeck:1998az}. The constraints 
are then referred to as the $\lambda$-variables.

The Z4 formulation~\cite{Bona:2003fj,Bona:2004yp,Bona:2003qn,
Bona:2004ky,Bona:2009ee,Bona:2010wn} has both propagating constraints 
and admits a constraint damping scheme~\cite{Gundlach:2005eh}. The Z4 
formulation has a close relationship with the generalized 
harmonic formulation, and the damping scheme is essentially the same 
for both systems.
The damping scheme was a crucial ingredient in the 
first successful evolution of orbiting binary black holes through 
merger~\cite{Pretorius:2005gq}. Analytic calculations demonstrating 
that constraint violations will be damped away are performed in the 
frozen coefficient approximation. On the basis of these calculations 
one expects that the damping scheme will be effective, 
in numerical applications, on constraint violations that are of 
low amplitude and high frequency in spacetimes that are close to 
stationarity. Since the constraint damping scheme is a modification 
of the continuum equations, this high-frequency should be resolved 
on the numerical mesh. It is not clear what effect the damping 
scheme will have on ill-resolved numerical noise.

A conformal decomposition of Z4, called Z4c, was 
proposed~\cite{Bernuzzi:2009ex,Ruiz:2010qj} with the hope of bringing 
the advantages of propagating constraints and the constraint damping 
scheme to the puncture method~\cite{Campanelli:2005dd,Baker:2005vv} for 
the evolution of binary black holes. Here we continue that investigation, 
considering more carefully the effect of the constraint damping scheme 
on numerical evolutions. We address the following questions: (i).~Under 
what conditions can the theoretically predicted damping rates be recovered
in the numerical approximation? (ii).~How effective is the damping scheme 
in astrophysically relevant spacetimes? (iii).~In practical applications 
what are reasonable values for the constraint damping coefficients? 

A variation of the conformal decomposition has been recently presented
in~\cite{Alic:2011gg}.
There it was found that the constraint damping terms are 
essential for stable long-term 3D evolutions of binary black holes and the
gauge wave test. Since the Z4c conformal 
decomposition differs from that in~\cite{Alic:2011gg} by nonprincipal 
terms and implementation details~(e.g.~constraints projection and
summation-by-part operators), our study does not necessarily apply in
that case. More work is required to carefully evaluate the role of the
constraint damping scheme in that case.

Because of the obvious computational overhead of working in three 
dimensions, we once again present numerical results in spherical symmetry. 
Note that since the constraint damping scheme is a modification to the 
continuum Z4c formulation, our results are expected to reflect the 
behavior of the full system. Working in spherical symmetry furthermore 
affords us the possibility of performing a thorough study in the parameter 
space of constraint damping coefficients. 

In Sec.~\ref{sec:Theory} we summarize the equations of motion
for the Z4c formulation and describe the constraint damping scheme 
we employ. We also present the expected theoretical rates of damping 
in the high-frequency, frozen coefficient approximation. In 
Sec.~\ref{sec:Numerics} we present our numerical study.
Finally we conclude in Sec.~\ref{sec:Conclusion}.

\paragraph*{Notation.} Geometric units are employed. 
Standard notation for the 3+1 general relativity is used, e.g.  
$\partial_i$, partial derivative with respect to 
coordinates~$x^i$, $i=1,2,3$, $D_i$, 3-covariant derivative, 
$\Lie_\beta$, Lie derivative along the vector $\beta^i$,
$\gamma$, determinant of the 3-metric $\gamma_{ij}$,
$K_{ij}, \, \alpha, \, \beta^i$, extrinsic curvature, lapse 
function and shift vector. In the perturbed flat-space 
simulations there is no natural scale of time; there we give 
the time in abitrary units.

\section{The Z4c constraint damping scheme}
\label{sec:Theory}

\subsection{The Z4c formulation}
\label{sec:Z4c}

In $3+1$ form the field equations of the Z4c formulation of  
general relativity for the three-metric and extrinsic curvature 
read  
\begin{align}
\p_t\gamma_{ij}=&-2\alpha K_{ij}+\Lie_\beta\gamma_{ij},\\
\p_tK_{ij}=&-D_iD_j\alpha+\alpha[R_{ij}-2K_i{}^kK_{kj}+KK_{ij}\nonumber\\
&+2\hat{D}_{(i}Z_{j)}-\kappa_1(1+\kappa_2)\gamma_{ij}\Theta]\nonumber\\
&+4\pi\alpha[\gamma_{ij}(S-\rho_{{\textrm {\tiny ADM}}})-2 S_{ij}]
+\Lie_\beta K_{ij}, 
\end{align}
where we use the notation
\begin{align}
\hat{D}_{i}Z_{j}&\equiv
\gamma^{-\frac{1}{3}}\gamma_{kj}\p_{i}[\gamma^{\frac{1}{3}}Z^k].
\end{align}
The constraints~$\Theta,\, Z_i$ evolve according to 
\begin{align}
\label{eqn:Theta_dot}
\p_t\Theta=&\alpha[\frac{1}{2} H + \hat{D}^iZ_i
-\kappa_1(2+\kappa_2)\Theta]+\Lie_\beta\Theta,\\
\label{eqn:Z_dot}
\p_tZ_i=& \alpha[M_i+D_i\Theta-\kappa_1Z_i]
+\gamma^{\frac{1}{3}}Z^j\p_t[\gamma^{-\frac{1}{3}}\gamma_{ij}]+\beta^j\hat{D}_jZ_i,
\end{align}
where the Hamiltonian and momentum constraints $H,\, M_i$ are 
given by
\begin{align}
H=&R-K_{ij}K^{ij}+K^2 - 16\pi\rho_{{\textrm {\tiny ADM}}}=0,\\
M_i=&D^j[K_{ij}-\gamma_{ij}K]-8\pi S_i=0.
\end{align}
Their time dependence can be computed as (neglecting matter terms), 
\begin{align}
\label{eqn:Ham_dot}
\p_tH=&-2\alpha D^iM_i-4M_iD^i\alpha+2\alpha K H\nonumber\\
&+ 2\alpha\left(2K\gamma^{ij}-K^{ij}\right)\hat{D}_{(i}Z_{j)}\nonumber\\
&- 4\kappa_1(1+\kappa_2)\alpha K \Theta+\Lie_\beta H,\\
\label{eqn:Mom_dot}
\p_tM_i=&-\half\alpha D_i H + \alpha K M_i -(D_i\alpha) H \nonumber\\
&+D^j\left(2\alpha\hat{D}_{(i}Z_{j)}\right)-D_i\left(2\alpha\gamma^{kl}\hat{D}_{(k}Z_{l)}\right)\nonumber\\
&+2\kappa_1(1+\kappa_2)D_i(\alpha\Theta)+\Lie_\beta M_i.  
\end{align}
From Eq.~(\ref{eqn:Theta_dot}-\ref{eqn:Z_dot}) one sees that
$\Theta,\, Z_i$ behave as $\lambda$ variables if the free parameters
$\kappa_{1,2}$ are properly chosen. Consequently the Einstein constraint 
are damped for $\kappa_{1}>0$ and $\kappa_2>-1$~\cite{Gundlach:2005eh}. The 
constraint subsystem is closed. If the constraints~$\Theta,Z_i$ and 
$H, M_i$ are satisfied in one hypersurface they will remain satisfied 
at all times. Introducing the following variables and definitions,
\begin{align}
&\tilde{\gamma}_{ij} = \gamma^{-\frac{1}{3}}\gamma_{ij},\quad
\chi = \gamma^{-\frac{1}{3}}, \\
&\hat{K} = \gamma^{ij}K_{ij} - 2\Theta\ ,\quad
\tilde{A}_{ij}=\gamma^{-\frac{1}{3}}(K_{ij}-\frac{1}{3}\gamma_{ij}K), \\
&\tilde{\Gamma}^{i} = 2 \tilde{\gamma}^{ij} Z_j + \tilde{\gamma}^{ij}
  \tilde{\gamma}^{kl}\tilde{\gamma}_{jk,l},\quad
\tilde{\Gamma}_{\textrm{d}}{}^i
=\tilde{\gamma}^{jk}\tilde{\Gamma}^{i}{}_{jk}, 
\end{align}
the conformal evolutions equations, Z4c, read,
\begin{align}
\p_t \chi =& \frac{2}{3}\chi[\alpha(\hat{K}+2\Theta) - D_i\beta^i],\\
\p_t \tilde{\gamma}_{ij} =& -2\alpha\tilde{A}_{ij}+\beta^k
\tilde{\gamma}_{ij,k}+2\tilde{\gamma}_{k(i}\beta^k_{,j)}-\frac{2}{3}
\tilde{\gamma}_{ij}\beta^k_{,k}\ ,\\
\p_t \hat{K}    =& -D^iD_i\alpha + \alpha[\tilde{A}_{ij}\tilde{A}^{ij}
+\frac{1}{3}(\hat{K}+2\Theta)^2]\nonumber\\
&+4\pi\alpha[S+\rho_{{\textrm {\tiny ADM}}}]
+\alpha\kappa_1(1-\kappa_2)\Theta+\beta^i\hat{K}_{,i}\\
\p_t \tilde{A}_{ij} =& \chi[-D_iD_j\alpha
+\alpha (R_{ij}-8\pi S_{ij})]^{\textrm{tf}}\nonumber\\
& +\alpha[(\hat{K}+2\Theta)\tilde{A}_{ij} - 2\tilde{A}^k{}_i\tilde{A}_{kj}]
\nonumber\\
& + \beta^k\tilde{A}_{ij,k} +2\tilde{A}_{k(i}\beta^k_{,j)}-\frac{2}{3}\tilde{A}_{ij}
\beta^{k}{}_{,k} \\
\p_t \tilde{\Gamma}^{i} =& -2\tilde{A}^{ij}\alpha_{,j}+2\alpha
[\tilde{\Gamma}^i_{jk}\tilde{A}^{jk}-\frac{3}{2}\tilde{A}^{ij}\ln(\chi)_{,j}
\nonumber\\
&-\frac{1}{3}\tilde{\gamma}^{ij}(2\hat{K}+\Theta)_{,j}
-8\pi\tilde{\gamma}^{ij}S_j]+\tilde{\gamma}^{jk}\beta^i_{,jk}\nonumber\\
&
+\frac{1}{3}\tilde{\gamma}
^{ij}\beta^k_{,kj}+\beta^j\tilde{\Gamma}^i_{,j}
-\tilde{\Gamma}_{\textrm{d}}{}^j\beta^i_{,j}+\frac{2}{3}
\tilde{\Gamma}_{\textrm{d}}{}^i\beta^j_{,j}\nonumber\\
&-2\alpha\kappa_1(\tilde{\Gamma}^i
-\tilde{\Gamma}_{\textrm{d}}{}^i), \\
\p_t\Theta =&\alpha[\frac{1}{2} R - \frac{1}{2} \tilde{A}_{ij}\tilde{A}^{ij} +\frac{1}{3}(\hat{K}+2\Theta)^2 \nonumber\\
           &-8\pi\rho_{\mathrm{ADM}} - \kappa_1(2+\kappa_2)\Theta]+ \Lie_\beta\Theta.
\end{align}
Here the intrinsic curvature is written as
\begin{align}
R_{ij} =& R^{\chi}{}_{ij} + \tilde{R}_{ij},\\
\tilde{R}^{\chi}{}_{ij} =&
\frac{1}{2\chi}\tilde{D}_i\tilde{D}_j\chi+\frac{1}{2\chi}
\tilde{\gamma}_{ij}\tilde{D}^l\tilde{D}_l\chi\nonumber\\
&-\frac{1}{4\chi^2}\tilde{D}_i\chi\tilde{D}_j\chi-\frac{3}{4\chi^2}
\tilde{\gamma}_{ij}\tilde{D}^l\chi\tilde{D}_l\chi,\\
\tilde{R}_{ij} =&
 - \frac{1}{2}\tilde{\gamma}^{lm}
\tilde{\gamma}_{ij,lm} +\tilde{\gamma}_{k(i|}\tilde{\Gamma}
^k_{|,j)}+\tilde{\Gamma}_{\textrm{d}}{}^k
\tilde{\Gamma}_{(ij)k}\nonumber\\
&+\tilde{\gamma}^{lm}\left(2\tilde{\Gamma}^k_
{l(i}\tilde{\Gamma}_{j)km}+\tilde{\Gamma}^k_{im}
\tilde{\Gamma}_{klj}\right).
\end{align}
The equations above are constrained by two algebraic expressions, 
$\ln(\det\tilde{\gamma})=0$ and $\tilde{\gamma}^{ij}\tilde{A}_{ij}=0$, 
which are explicitly imposed during the numerical evolution. A 
solution of the evolution system is a solution of the Einstein 
system provided that~$\Theta$, $Z_i$, $H$ and $M_i$ vanish. In 
our numerical applications we close the system with the puncture 
gauge choice~\cite{Bona:1994dr,Alcubierre:2002kk},
\begin{align}
\p_t\alpha=& -\mu_L\alpha^2\hat{K}+\Lie_\beta\alpha,\\
\p_t\beta^i=& \mu_S\alpha^2\tilde{\Gamma}^i  - \eta \beta^i 
+ \beta^j\p_j\beta^i, 
\end{align}
with $\mu_L=2/\alpha$, $\mu_S=1/\alpha^2$ and $\eta=2/M$, where 
$M$ is the ADM mass of the spacetime. Unless stated otherwise, we employ 
the constraint-preserving boundary condition of~\cite{Ruiz:2010qj}.

For a more thorough introduction to the Z4c 
formulation we refer the reader to~\cite{Bernuzzi:2009ex,Ruiz:2010qj}. 
The full formulation generically forms a strongly hyperbolic 
system of partial differential equations, except in special cases 
which we do not discuss here. 

\subsection{Theoretical damping rates}
\label{sec:Plane-wave}

Here we briefly review the results of~\cite{Gundlach:2005eh}, 
see also~\cite{Friedrich:2005pd} for a discussion of stability of the 
undamped nonlinear constraint system. Consider the 
subsystem~(\ref{eqn:Theta_dot}-\ref{eqn:Z_dot},
\ref{eqn:Ham_dot}-\ref{eqn:Mom_dot}) when the initial data 
is constraint violating. 
We consider a small constraint-violating
perturbation on a background which satisfies the Z4c equations 
of motion. Working in the frozen coefficient approximation we 
can choose coordinates at a point so that the spatial metric 
is just that of flat-space, and the lapse is unity. The only 
nontrivial component of the metric is the shift, which can 
only be made to vanish if we allow ourselves the freedom to 
change the spatial slice, which we take as 
given~\cite{Ruiz:2007hg}. We discard nonprincipal terms 
involving products of the perturbation and the background 
curvature, extrinsic curvature, and matter sources, which is 
inconsistent with our first-order perturbation approach. 
The inclusion of matter sources without background curvature 
terms should give the correct damping rates for test matter fields 
on flat space. Such an analysis is not expected to give the 
rates for compact stars as evolved in Sec.~\ref{sec:Numerics}; 
nontrivial background curvature, extrinsic curvature and matter 
source terms will affect the damping rates, but for a consistent 
analysis all of these effects should be considered together. 
Already in the variable coefficient approximation such an 
analysis will be challenging. We will use the shift only to 
illustrate that it does not play any role on the damping rates. 
Besides the inclusion of the shift our calculations are exactly 
the same as those of~\cite{Gundlach:2005eh}. To start with, we 
make a plane wave ansatz 
\begin{align}
\Theta=e^{st + i\omega_ix^i}\hat{\Theta},&\qquad
Z_i=e^{st + i\omega_ix^i}\hat{Z}_i,
\end{align}
for the solution of the constraint subsystem, with complex $s$ 
and real $\omega_i$. We restrict to the special case~$\kappa_2=0$, 
and write $\kappa_1=k$. Rewriting the constraint subsystem in fully 
second order in time form, we find that following eigenvalue 
problem 
\begin{align}
\left(\begin{array}{cc}
  \lambda
+ 2s k+2i\omega\beta
  k & i\omega k \\
     0 & \lambda
+s k-i\omega\beta k
\end{array}\right)
\left(\begin{array}{c}
\hat{\Theta}\\
\hat{Z}_{\hat\omega}
\end{array}\right)&=0,\\
\left(\begin{array}{c}
 \lambda +s k-i\omega k\beta
\end{array}\right)
\left(\begin{array}{c}
\hat{Z}_{A}
\end{array}\right)
&=0,
\end{align}
must be satisfied, where we write 
\begin{align}
\lambda&=s^2+\omega^2-2 i\omega s\beta 
- \omega^2\beta^2,
\end{align}
and $\hat{Z}_{\hat\omega}$ stands for the component of $\hat{Z}_i$ in the
$\hat{\omega}^i$ direction, while $\hat{Z}_{A}$ are the components transverse 
to the unit wave vector~$\hat{\omega}^i$ and $\beta=\beta^i\hat{\omega}_i$. The 
symbol of this equation generically has a complete set of eigenvectors, so the 
system can be rotated to diagonal form,
\begin{align}
\left(\begin{array}{cc}
\lambda_\Theta & 0\\
0 & \lambda_{Z_{\hat{\omega}}}
\end{array}\right)
\left(\begin{array}{c}
s\hat{\Theta}+i\omega(\hat{Z}_{\hat{\omega}}-\beta\hat{\Theta})\\
\hat{Z}_{\hat{\omega}}
\end{array}\right)
&=0,\\
\left(\begin{array}{c}
 \lambda_{Z_T} 
\end{array}\right)
\left(\begin{array}{c}
\hat{Z}_{A}
\end{array}\right)&=0.
\end{align}
In one dimension, identifying the direction $r$, one expects that the 
behavior of the combinations of primitive variables,
\begin{align}
u_\Theta&=\partial_t\Theta+\partial_rZ_r,\\
u_\omega&=Z_r
\end{align}
will be determined, in the linear regime, by the eigenvalues of the symbol 
provided that the shift is small. The eigenvalues are 
\begin{align}
\lambda_\Theta=& \lambda + 2s k-2i k\omega\beta,\nonumber\\
  s=&-k+i\omega\beta\pm\sqrt{k^2-\omega^2},
\end{align}
and
\begin{align}
\lambda_{Z_s} =& \lambda+s k-i\omega k\beta,\nonumber\\
s=&-\frac{k}{2}+i\omega\beta\pm\sqrt{\bigg(\frac{k}{2}\bigg)^2-\omega^2},
\end{align}
and finally
\begin{align}
\lambda_{Z_T}=& \lambda+s k-i\omega k\beta,\nonumber\\
 s=&-\frac{k}{2}+i\omega\beta\pm\sqrt{\bigg(\frac{k}{2}\bigg)^2-\omega^2}.
\end{align}
In the low-frequency limit $\omega\ll k$ we find 
that 
\begin{align}
s&\simeq -2 k+i\omega\beta,&\quad
s\simeq i\omega\beta-\frac{\omega^2}{2 k},
\nonumber\\
s&\simeq-k+i\omega\beta+\frac{\omega^2}{k},&\quad
s\simeq i\omega\beta-\frac{\omega^2}{k},
\end{align}
whereas in the high-frequency limit $k\ll\omega$, we have 
\begin{align}
s&\simeq -k+i\omega(\beta\mp 1),
&\quad s\simeq -\frac{k}{2}+i\omega(\beta\pm 1),
\end{align}
so at lower frequencies half of the modes are damped less.
In the high-frequency limit, the damping scheme causes 
a exponential decay of the constraints with a decay rate of 
$-k$ and $-k/2$ respectively.

\section{Numerical Results}
\label{sec:Numerics}

In this section we present our numerical results. Spherical 
symmetry is assumed. We perform a detailed analysis of the 
damping scheme applied to the evolution of flat spacetime with 
different constraint-violating perturbations in order to reproduce 
the analytic results and explore the regime not accessible to 
a pen-and-paper analysis. We consider then nontrivial initial 
data composed of either punctures or a compact star and evolve 
them using different values for the damping parameters. In this 
case the performance of the damping scheme in a strong-field 
regime on constraint violations related, essentially, to different 
truncation errors are investigated. The code employed is described 
in detail in~\cite{Bernuzzi:2009ex}. 
\paragraph*{Numerical setup.} 
In all numerical simluations we use fourth-order finite
differences for the discretization of the spatial derivatives of the metric fields.
For the time integration we use Runge-Kutta fourth order in
the vacuum and puncture tests. In the simulation of the compact
star we employ the Runge-Kutta third order in combination with a
high-resolution shock capturing based on the local Lax-Friedrichs flux and the 
convex-essentially-non-oscillatory interpolation for the reconstruction of the matter fields.
Table~\ref{tab:num_settings} summarizes the numerical settings employed for the results presented;
convergence tests were run on some cases (see also discussions in the following paragraphs).

\begin{table}[t]
  \caption{ \label{tab:num_settings} Setting used for the numerical
    simulations presented in this paper. The resolution is given in grid
    points, $r_\mathrm{\tiny out}$ is the coordinate location of the
    computational boundary, and $\alpha_{\rm CFL}$ is the
    Courant-Friedrichs-Levy factor used in the time-stepping.} 
  \begin{tabular}{|c|c|c|c|}\hline
    Initial data & Resolution & $r_\mathrm{\tiny out}$ & $\alpha_{\rm CFL}$\\ 
    \hline
    Flat        & 4000 & 100 a.u.& 0.5 \\
    Puncture    & 1000 & 50 M    & 0.5\\
    Star        & 2000 & 50 M    & 0.4\\
    \hline
 \end{tabular}
\end{table}
 
\subsection{Perturbed flat space-time experiments}
\begin{figure*}[t]
 \centering
 \includegraphics[width=.44\textwidth]{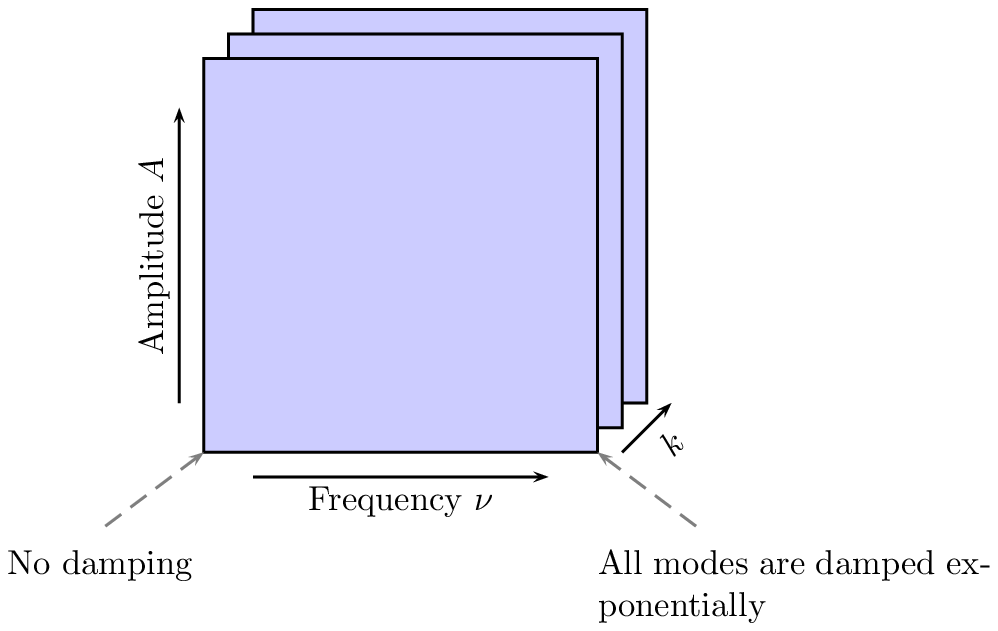}
 \includegraphics[width=.44\textwidth]{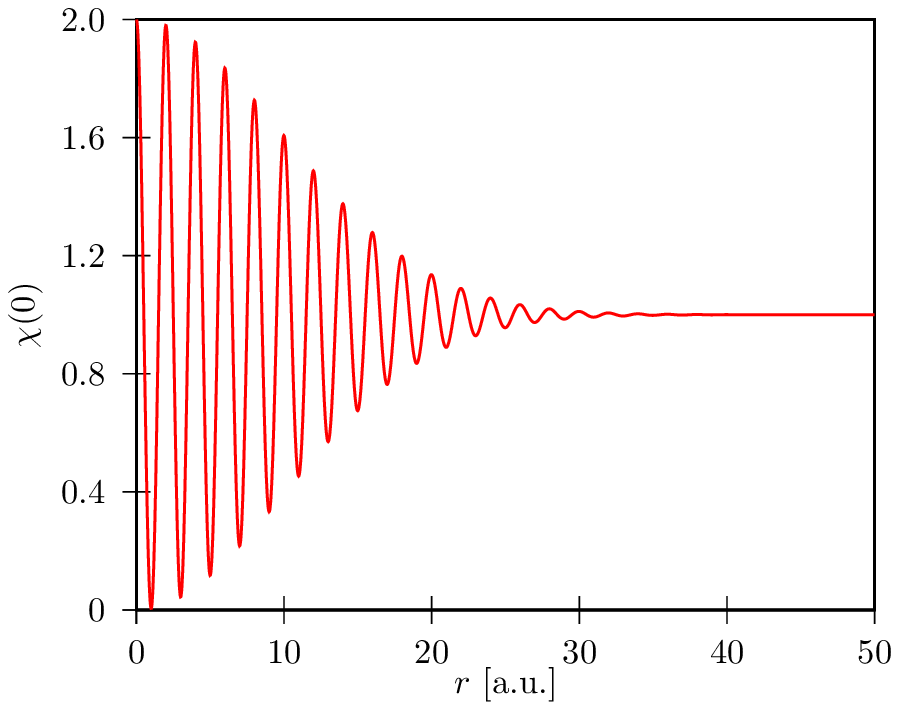}
 \caption{(Left) Parameter space for the constraint violation 
of flat space-time. The violation can be tuned by its
amplitude $A$ and frequency $\nu$. Analytically known are the low-frequency 
corner, where the violation is not damped and the 
high-frequency, low amplitude-corner where all modes are 
damped exponentially.
(Right) Shape of the constraint-violating initial data. A 
Gaussian curve with amplitude $A$ and full-width half-maximum 
$b$ is modulated with the frequency $\nu/b$. The parameter $\nu$ 
tells how many oscillations are within $b$. The figure shows the 
constraint violation in $\chi$ at $t=0$ with $A=1$, $b=10$ and $\nu=5$.} 
\label{fig:parspace}
\end{figure*}

\paragraph*{Initial data and parameter space.}
Perturbations of flat space-time are constructed, depending on which of
the two eigenmodes, $u_\Theta,u_\omega$, we want to analyze, by 
modifying the $\chi$ variable, 
\begin{equation}
\label{eq:id_chi}
  \chi(0,r)  =  1+ A
  \exp\left(-\frac{r^2}{2b^2}\right)\cos\left(\frac{2\pi\nu}{b}r\right),  
\end{equation}
or the $\tilde{\Gamma}^r$ variable,
\begin{equation}
\label{eq:id_gam}
  \tilde{\Gamma}^r(0,r) = A r
  \exp\left(-\frac{r^2}{2b^2}\right)\cos\left(\frac{2\pi\nu}{b}r\right)
  \ . \\  
\end{equation}
Simulations employing the first kind of initial data are analyzed by
looking at the eigenmode $u_\Theta$, while those employing the second
kind by looking at $u_\omega$. The eigenmodes are Fourier-transformed 
in space for every time step, and their decay is studied by means of the 
power spectral density (PSD) at the frequency $\tnu=\frac{\nu}{b}$.

The parameter space, depicted in Fig.~\ref{fig:parspace} (left panel), 
is spanned by the amplitude $A$ and the frequency $\nu$ of the initial
constraint violation. We vary also $\kappa_1=k\in[0,1]$ but keep for
simplicity $\kappa_2=0$. The parameter $b=10$ (fixed) introduces a
length scale to the problem, which is useful for tuning the 
frequency~$\nu$ (number of cycles in the period $b$, see right panel 
of Fig.~\ref{fig:parspace}). To evaluate the strength of the 
perturbation, the value of the perturbation's amplitude~$A$ should be 
compared with unity, see Eq.~(\ref{eq:id_chi}). 

\begin{figure*}[t]
 \centering
 \includegraphics[width=.44\textwidth]{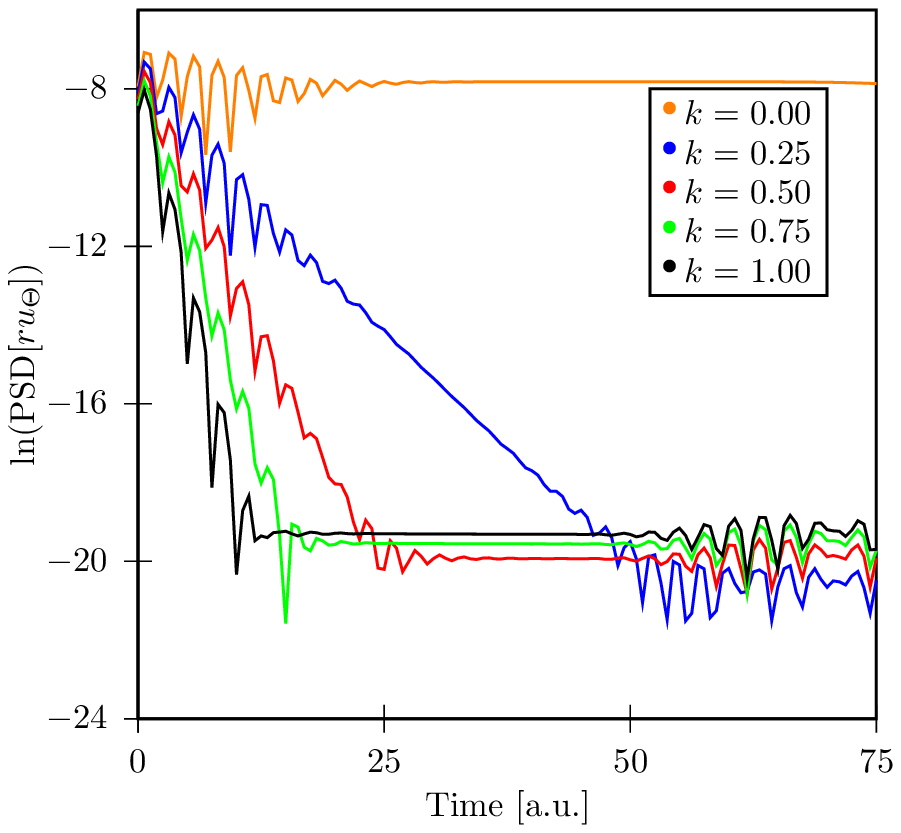}
 \includegraphics[width=.44\textwidth]{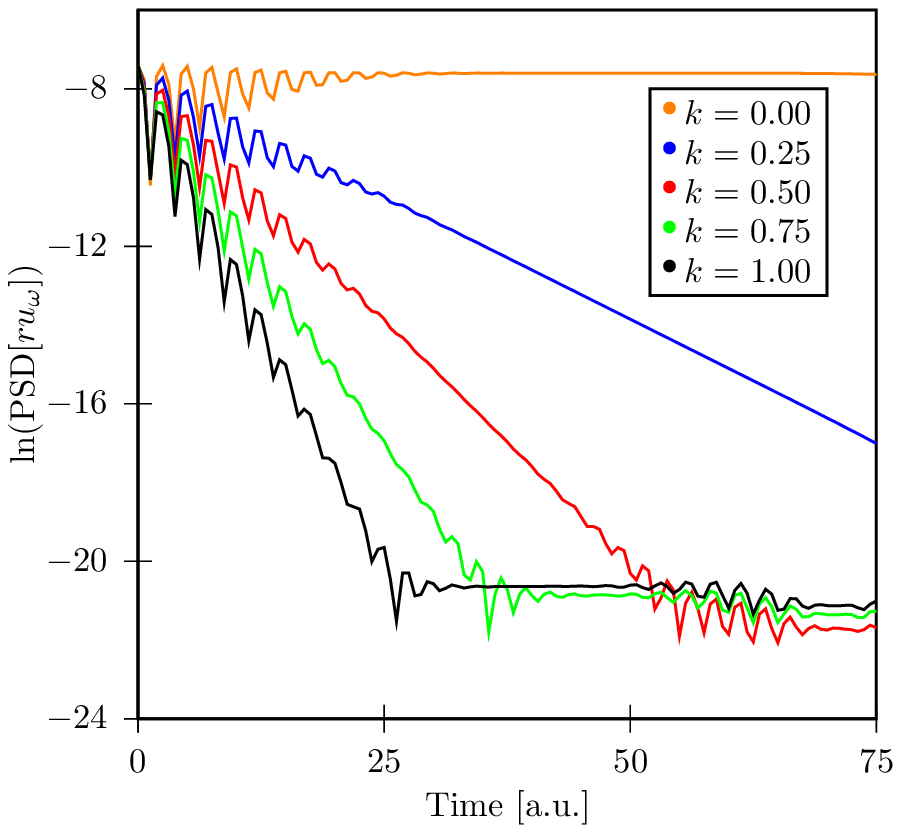}
 \caption{Behavior of the eigenmodes $u_\Theta$ (left) and $u_\omega$ 
          (right) for high-frequency $\nu=10$, low-amplitude $A=10^{-4}$ 
          constraint violation. The modes are extracted in the Fourier 
          space at $\bar\nu=1$. For no damping 
          ($k=0$) the modes are stay constant. For a $k>0$ the modes are damped 
          exponentially with different damping rates.}
 \label{fig:hfla:decay}
\end{figure*}

\begin{figure*}[t]
  \centering
  \includegraphics[width=.44\textwidth]{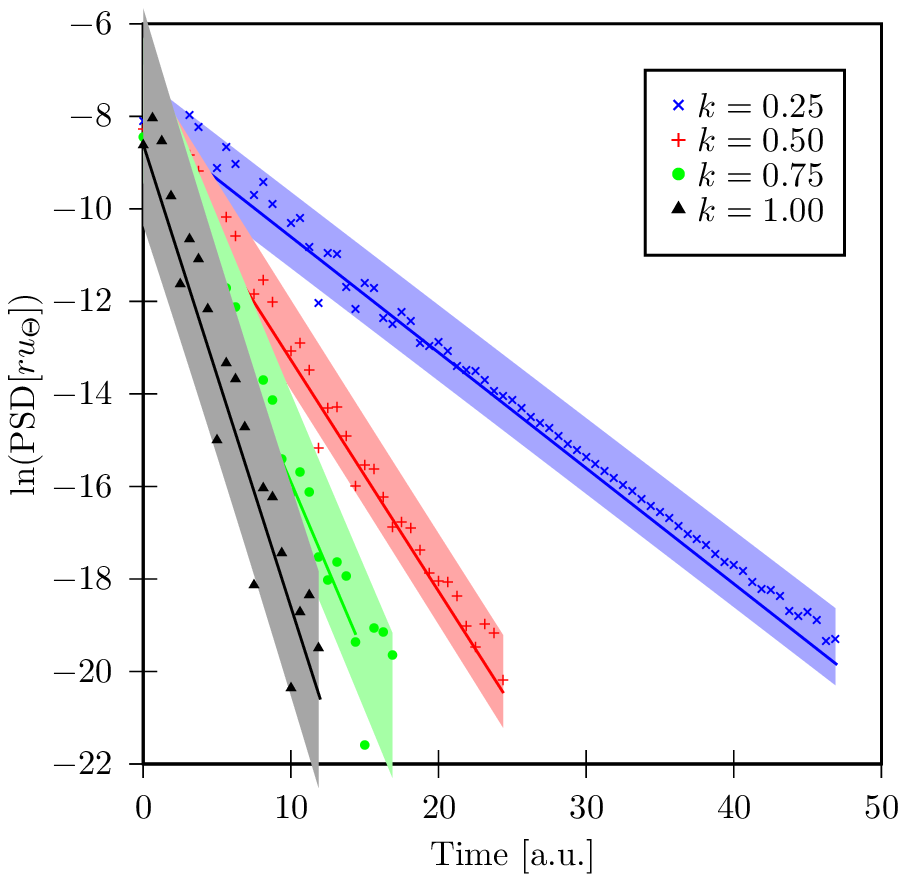}
  \includegraphics[width=.44\textwidth]{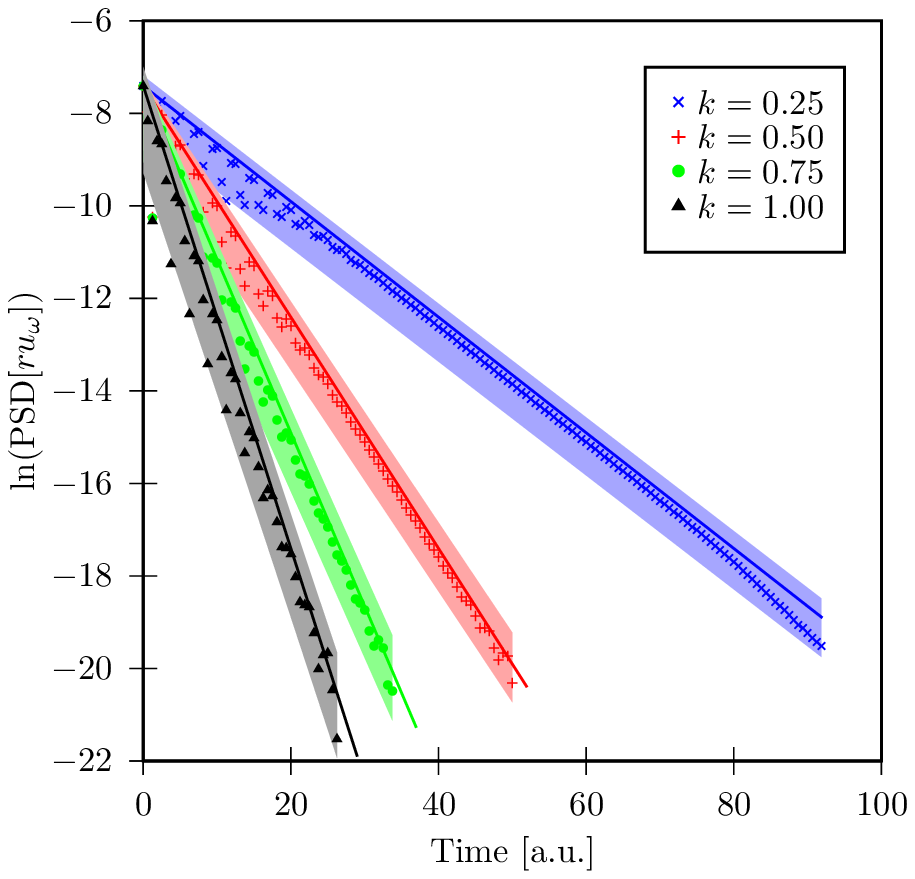}
 \caption{Fit of the exponential decay of the eigenmodes $u_\Theta$ (left) 
           and $u_\omega$ (right) for high-frequency $\nu=10$ and low-amplitude 
           $A=10^{-4}$ constraint violation. The analytically 
           predicted decay rates (solid lines) lie in every case within 
           the $68\%$ confidence interval of the fits which are represented 
           in the figure by the shaded regions.}
 \label{fig:hfla:decay:fits}
\end{figure*}

\paragraph*{High-frequency, low-amplitude corner. }
In this region of the parameter space the analytical results hold. 
We set $\nu=10$ ($\tnu=1$) and $A=10^{-4}$, 
Fig.~\ref{fig:hfla:decay} display the results obtained.
The numerical data show an exponential decay of the PSD at the 
induced frequency $\tnu$ for both eigenmodes and for different 
values of the damping parameter $k$.  
The rate is quantified by linear fitting as displayed in
Fig.~\ref{fig:hfla:decay:fits}. 
Table~\ref{tab:hfla:decay} reports the decay rates for different
values of $k$ and different grid resolution ($n$ is the number of 
grid points),
together with the computed fitting error. 
Almost for every case the analytically predicted values,
i.e.~$s\approx k$ (for $u_\Theta$) and $s=\frac{k}{2}$ (for
$u_\omega$), lie within the error range of the numerically found number. 
For increasing resolution the error gets smaller. Note that the large error is caused 
by the oscillations of the modes. The decay rates of the modes agree much better with 
the analytically predicted values than our conservative error estimate suggests 
(see Fig.~\ref{fig:hfla:decay:fits})

\begin{table}[t]
  \centering 
  \caption{Fits results for the decay rates for different resolutions. 
    The parameters of the initial perturbation are $\nu=10$ and $A=10^{-4}$.}
  \label{tab:hfla:decay}
  \begin{tabular}[t]{|l||l|l|l|l|}
  \hline 
  $k$&$s$ ($n=2000$) & $s$ ($n=4000$)& $s$ ($n=8000$)& $s_{\rm analytic}$\\
  \hline
  $0.25$& $-0.21\pm0.02$ & $-0.24\pm0.02$ & $-0.25\pm0.01$& $-0.25$  \\
  $0.50$& $-0.49\pm0.06$ & $-0.51\pm0.04$ & $-0.50\pm0.04$& $-0.50$  \\
  $0.75$& $-0.74\pm0.09$ & $-0.78\pm0.08$ & $-0.77\pm0.05$& $-0.75$  \\
  $1.00$& $-0.93\pm0.16$ & $-1.02\pm0.20$ & $-1.02\pm0.14$& $-1.00$ \\     
  \hline    
  $0.25$& $-0.12\pm0.01$  & $-0.12\pm0.01$ & $-0.12\pm0.01$ & $-0.125$ \\
  $0.50$& $-0.24\pm0.02$  & $-0.24\pm0.02$ & $-0.24\pm0.01$ & $-0.250$\\
  $0.75$& $-0.35\pm0.04$  & $-0.35\pm0.04$ & $-0.36\pm0.02$ & $-0.375$\\
  $1.00$& $-0.47\pm0.07$  & $-0.48\pm0.07$ & $-0.48\pm0.03$ & $-0.500$\\    
  \hline
  \end{tabular}
\end{table}  

\begin{figure*}[t]
  \centering
  \includegraphics[width=.44\textwidth]{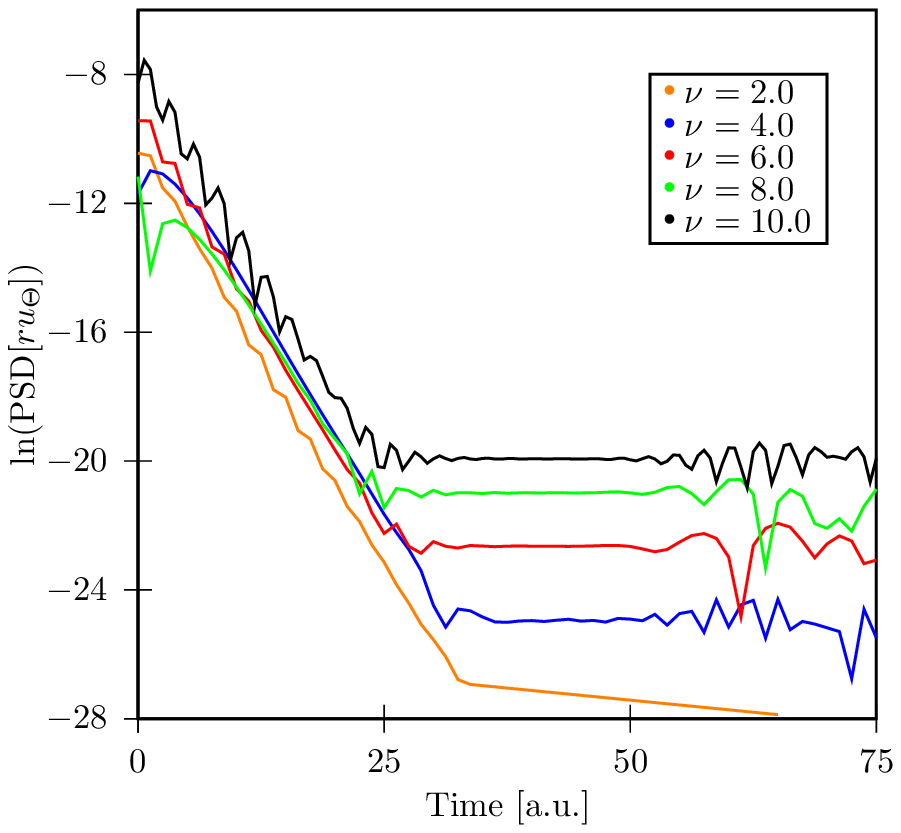}
  \includegraphics[width=.44\textwidth]{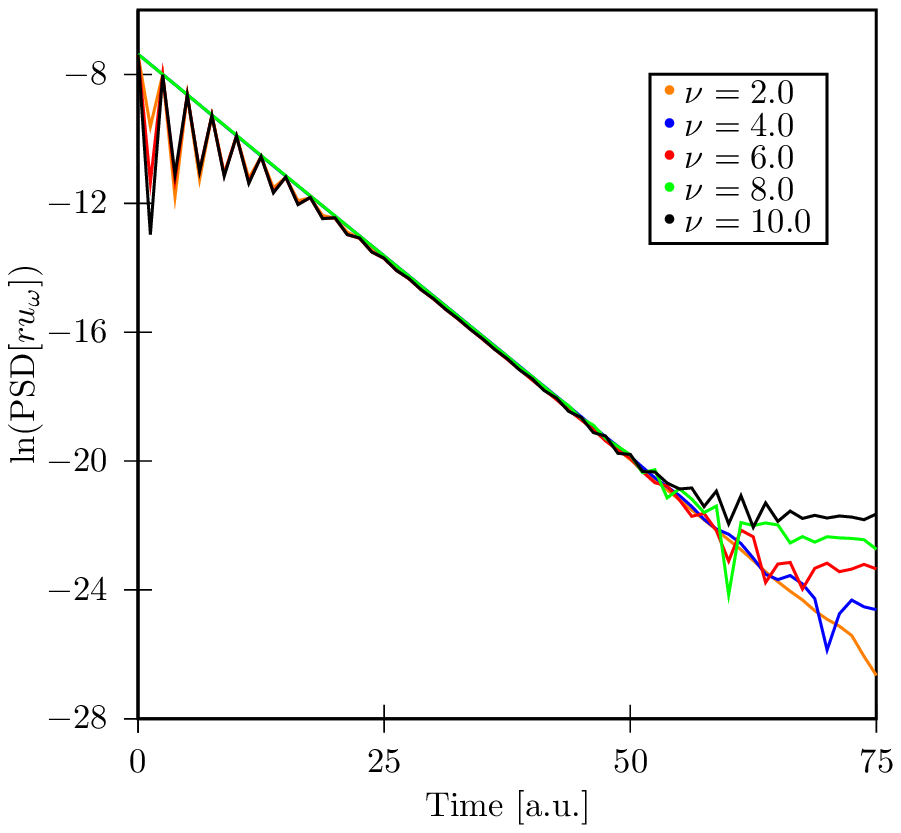}
  \caption{The rate of the exponential decay of the eigenmodes $u_\Theta$ (left) and $u_\omega$ (right) 
           for low-amplitude $A=10^{-4}$ constraint violations
           stays constant in the frequency range $\nu\in[2,10]$.}
  \label{fig:lfla:decay}
\end{figure*}

\paragraph*{From high to low frequency.} To explore the low-frequency regime, 
we keep the low amplitude $A=10^{-4}$ and vary the frequency in the
range $\nu=[0,10]$.
Analytic results indicates that there is no
damping for constant in space modes (zero-frequency modes). 
The numerical experiments show that, for decreasing values of the
initial perturbation frequency in the range $[2,10]$, the damping
scheme remains effective with the 
analytic exponential decay rates. The behavior is displayed in
Fig.~\ref{fig:lfla:decay}.  

\begin{figure*}[t]
  \centering
  \includegraphics[width=.44\textwidth]{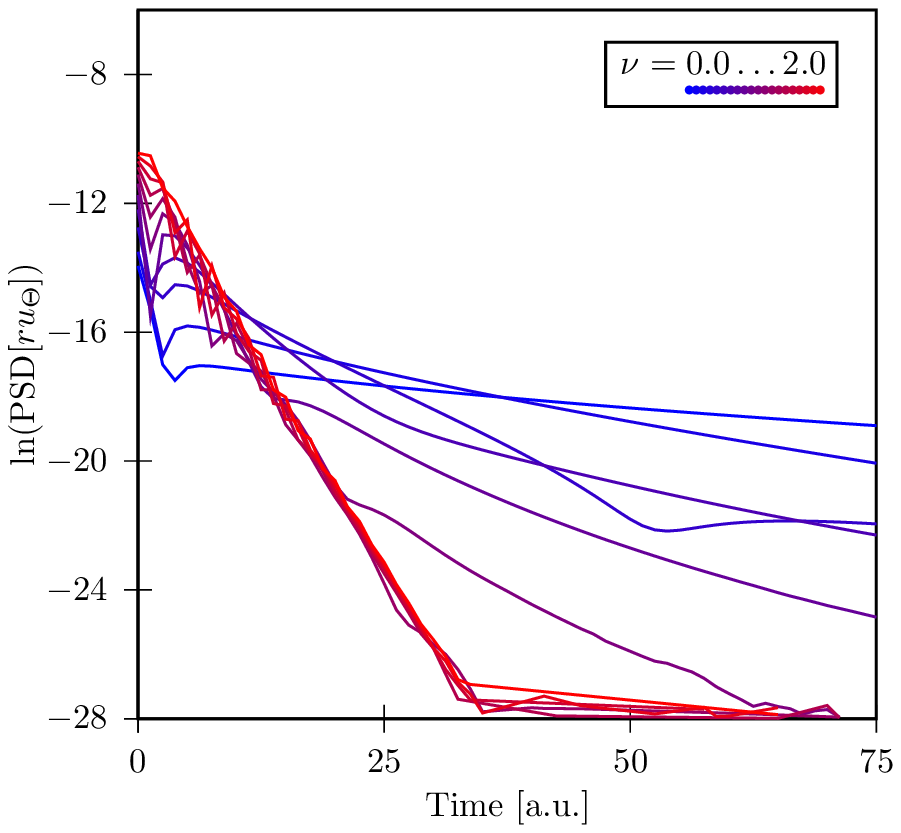}    
  \includegraphics[width=.44\textwidth]{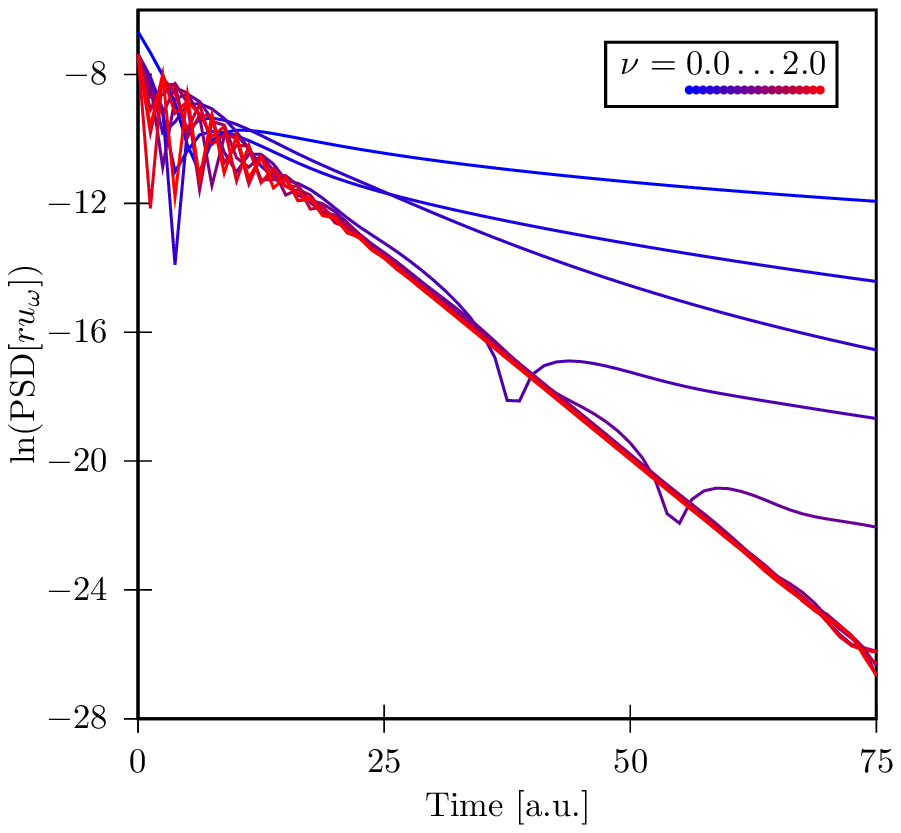}  
  \caption{In the small frequency-range $\nu\in[0,2]$ the transition between 
           exponential damping and no damping happens. The figure shows for 
           low-amplitude $A=10^{-4}$ constraint violations the decay of the
           eigenmodes $u_\Theta$ (left) and $u_\omega$ (right). }
  \label{fig:lfla:decay:smallf}
\end{figure*}

The transition from exponential damping to no damping happens after the
second octave $\nu\in[0,2]$, Fig.~\ref{fig:lfla:decay:smallf}. As
demonstrated by the plot the transition is smooth and quite rapid.  
The experimental fact, observed here, that constraint -violating modes
of ``almost all'' nonzero frequencies are killed by the damping
scheme can be important in numerical relativity simulation. 

\paragraph*{From low to high amplitude.} Increasing the amplitude of
the perturbation, i.e.~moving from a perturbative regime to a fully
nonlinear situation is a delicate procedure. Our results can be
summarized as follows. 

High-amplitude perturbations, up to $A\simeq0.1$, are damped and the 
damping rates unaffected. The use of progressively higher
amplitudes first modifies the damping rates (constraint-violating modes
are less damped) and secondly leads to unphysical results and code
failures. The maximum amplitude which can be reached without changing the
damping parameter $k$ depends on the frequency of the initial perturbation.
For low-frequency initial perturbations higher amplitudes are effectively 
damped by the damping scheme than for high frequencies. Furthermore, it 
is not true that increasing $k$ 
generically allows for higher amplitudes. The dependence on the 
damping parameter is not monotonic, the optimal value for this problem 
has been experimentally found to be $k=0.5$.

\begin{figure}[t]
  \centering
  \includegraphics[width=.44\textwidth]{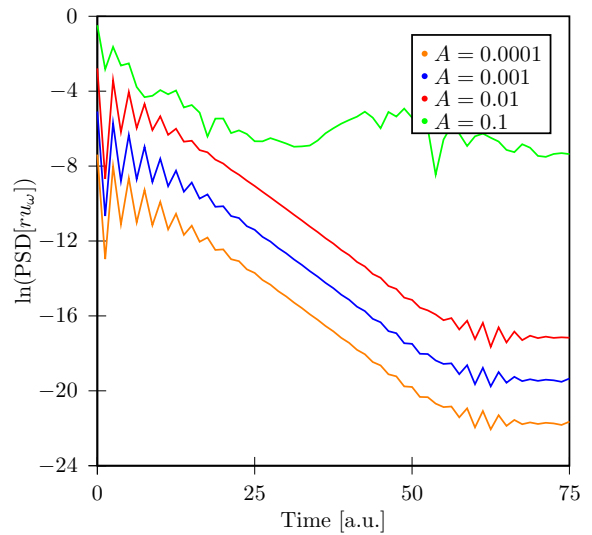}    
  \caption{Decay of the eigenmode $u_\omega$ for high-frequency constraint 
           violation with increasing amplitude $A$. Between $A=10^{-4}$ and $A=10^{-2}$
           the damping rate does not change. For very high-amplitude $A=10^{-1}$ the 
           damping of the eigenmode is not exponential anymore and the code does
           not give physically reasonable results.}
  \label{fig:hamp}
\end{figure}

Figure~\ref{fig:hamp} shows that, for $k=0.5$ and $\nu=10$, the
damping rates stay the same as in the low-amplitude case
for an amplitude range $A\in[0.0001,0.01]$.
Setting the amplitude to $A=0.1$, the damping is no longer exponential 
and, for higher values, the code gives no reasonable result.

\begin{figure}[t]
 \centering
 \includegraphics[width=.44\textwidth]{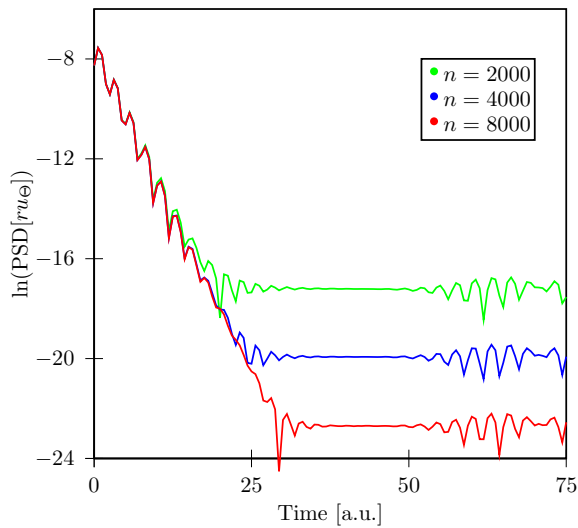}
 \caption{  \label{fig:resol} 
   The damping effect depends on the resolution of the initial
   constraint violation. The figure shows for high-frequency $\nu=10$, 
   low-amplitude $A=10^{-4}$ the damping of the eigenmode $u_\Theta$ for
   different resolutions. For high resolution the frequency is well-resolved 
   and therefore the damping effects lasts longer than for less well-resolved
   case.}
\end{figure}

\paragraph*{Resolution dependency.} 
Convergence of the results has been already reported in
Table~\ref{tab:hfla:decay} and  briefly discussed; we further comment
here focusing on representative simulations with $k=0.5$, $\nu=10$,
$A=10^{-4}$ with varying resolutions.  
As shown in Fig.~\ref{fig:resol} the damping effect holds longer for
higher resolutions,  i.e. if the frequency of the constraint violation
is better resolved, then the damping scheme works more effectively. 
This suggests that the damping scheme works as long as the 
frequency of the perturbation is well-resolved and it is partially
expected since it acts at the continuum level. 
The same effect could have been already anticipated from
Fig.~\ref{fig:lfla:decay}, which refers to a single resolution but
different frequencies which are resolved differently 
on the grid. 

\begin{figure}[t]
  \centering
  \includegraphics[width=.44\textwidth]{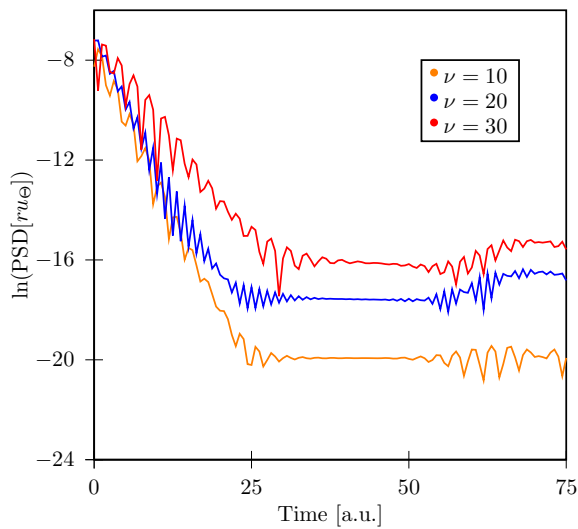}  
  \caption{ Increasing the frequency of the initial constraint violation
    with low-amplitude $A=10^{-4}$  even further to very high frequencies.
    At these high frequencies the constraint violation get less resolved
    which weakens the effect of the damping.
  } 
  \label{fig:vhf:decay} 
\end{figure}

\begin{figure*}[t]
  \centering
  \includegraphics[width=.44\textwidth]{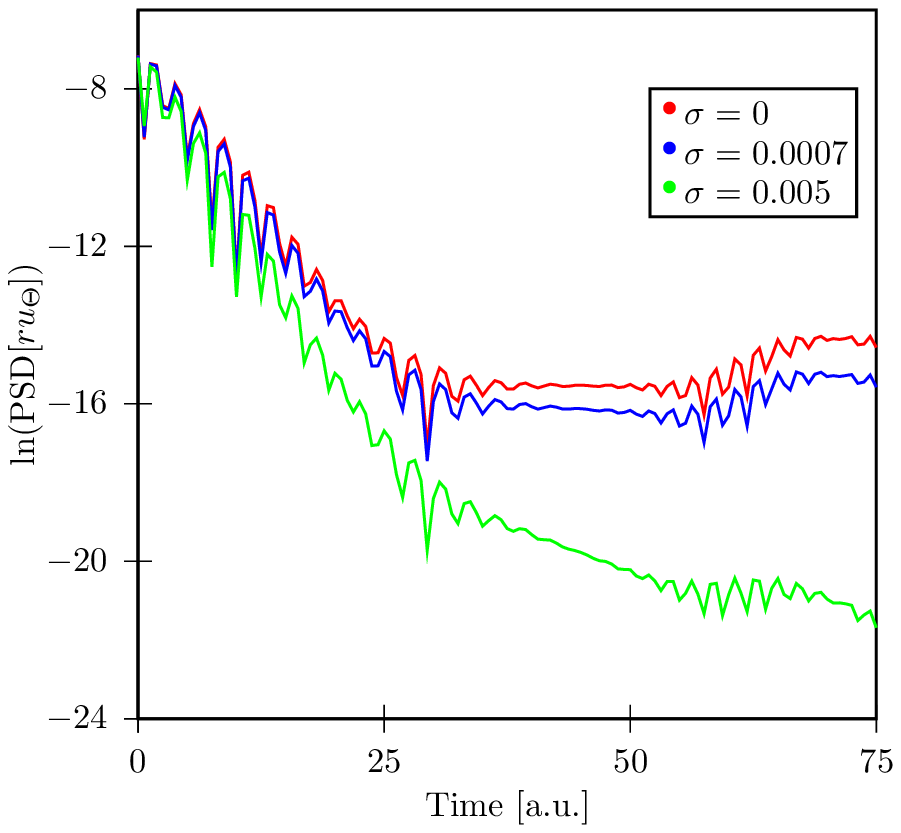}  
  \includegraphics[width=.44\textwidth]{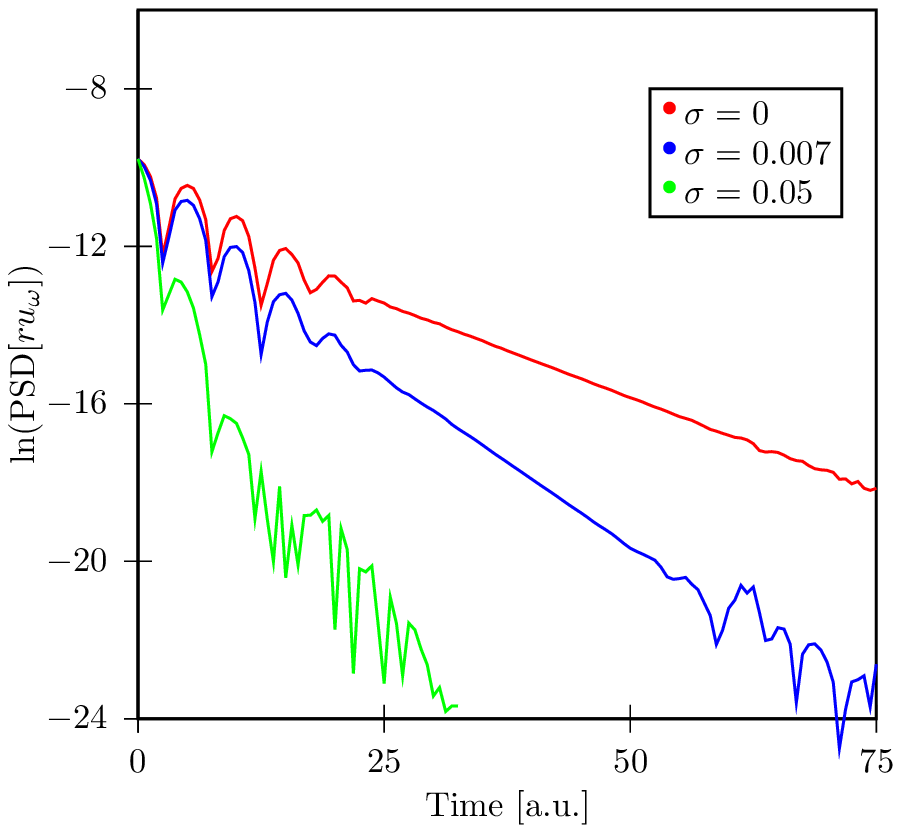}
  \caption{For initial constraint violations with low amplitude and very 
           high frequency, the artificial dissipation starts to 
           have an effect. The high-frequency modes are shifted by the 
           artificial dissipation to lower frequencies which are then damped.
           The figure shows the dependence of the decay of the
           eigenmode $u_\Theta$ on the artificial dissipation  
           parameter $\sigma$  using high-frequency $\nu=30$ (left)
           and random noise constraint violation initial data
           (right).} 
  \label{fig:vhf:diss} 
\end{figure*}

\paragraph*{Very high frequencies, grid modes and dependency on
  artificial dissipation.} 
The effect of the damping scheme on frequencies comparable to or of the
order of the numerical grid (grid modes) is finally  
studied~\footnote{ %
  We use now the term ``high'' not related to the scale $\bar{\nu}\gtrsim1$ of the
  continuum problem, i.e.~not for $\nu\sim b$,  
  but related to the characteristic frequency introduced by the
  numerical grid, i.e.~$\nu\sim 1/h$.}  
.   
For these tests we use $k=0.5$ and vary the frequency $\nu\in[10,30]$
(Note the grid spacing is $h=0.025$.). 
As demonstrated in Fig.~\ref{fig:vhf:decay}, 
if the frequency $\nu$ of the perturbation is increased further to a 
regime where the signal is not well-resolved, the damping becomes
progressively less  effective and deviates from the analytic
expectation. More importantly, the amount of artificial
dissipation~\cite{Gustafsson1995} plays a significant role. 
The use of the artificial dissipation filters out grid modes and
generically attenuates high frequencies which are aliased to lower
ones.  
Figure~\ref{fig:vhf:diss} (left panel) shows that the use of different amounts of
dissipation, $\sigma$, quantitatively changes the decay rate of the
eigenmodes. The higher $\sigma$ is, the higher the damping rate. However
one cannot expect to use arbitrary large values of $\sigma$, so a
balance between $k$ and $\sigma$ and the property of the solution have
to be studied case by case by performing convergence tests.  
In our test case we found that $\sigma=0.05$ roughly reproduce
the analytic damping rates.  

As a final test we evolve initial data containing random noise, 
\begin{equation}
  \chi(0,r) =  1+ A
  \exp\left(-\frac{r^2}{2b^2}\right)\mathrm{rand}([-1,1]) \ . 
\end{equation}
Note that, in principle, random noise differs from the high-frequency
perturbation used before because it has a flat spectrum. 
Figure~\ref{fig:vhf:diss} (right panel) shows that the effectiveness of the
damping scheme depends again strongly on the value of $\sigma$  
in the artificial dissipation operator used. However, differently from
the previous case, in this case we were not able to recover the
analytic damping rates 

\subsection{Punctures and compact star experiments}

\begin{figure*}
 \centering
 \includegraphics[width=.44\textwidth]{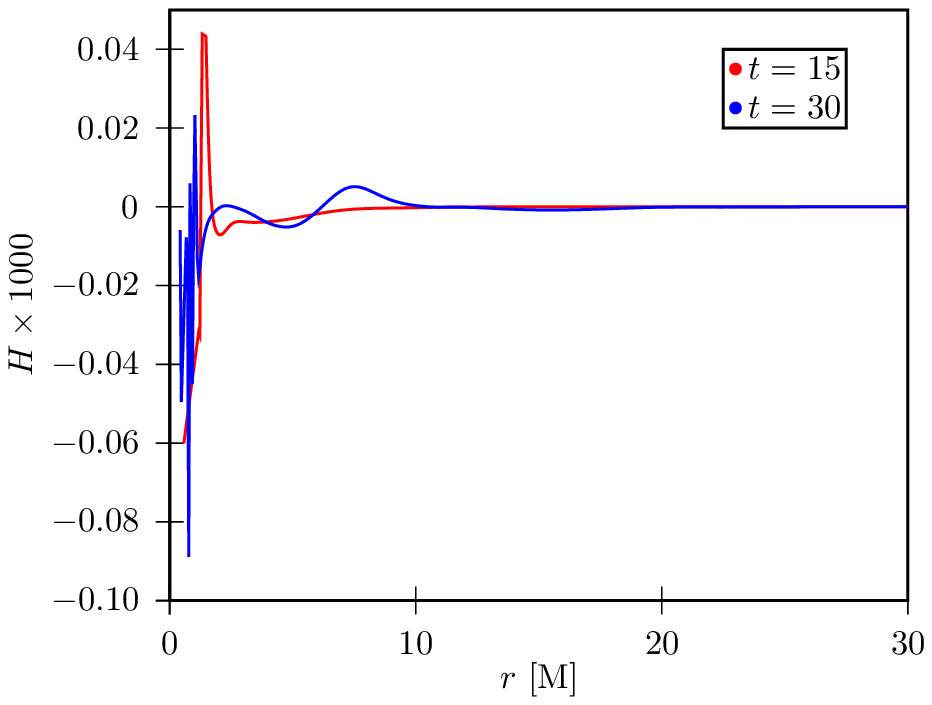}
 \includegraphics[width=.44\textwidth]{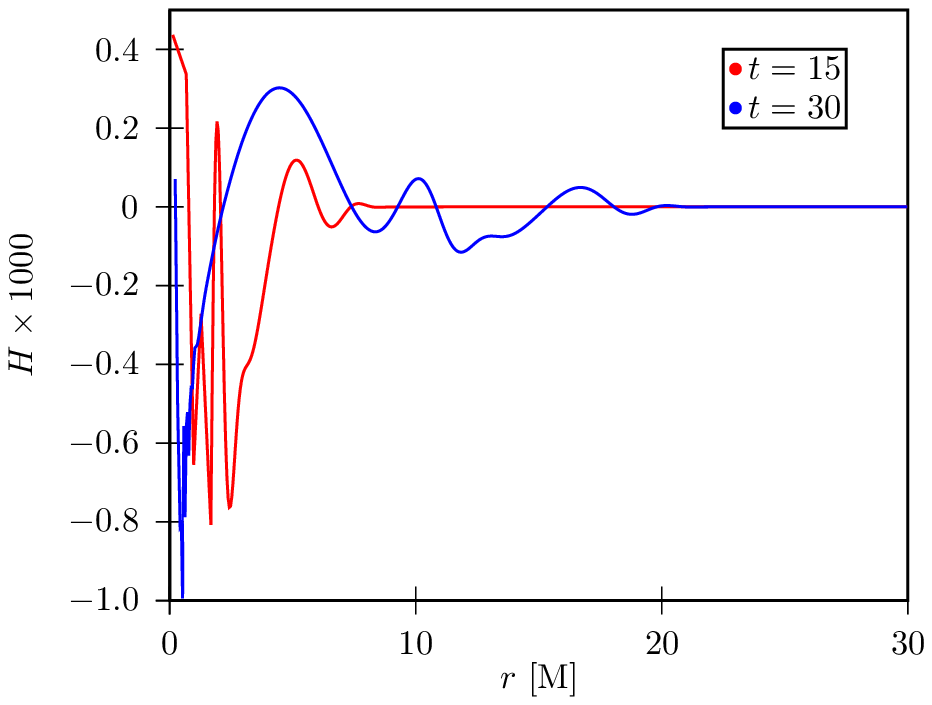}
 \caption{ \label{fig:punc:wave}  
   The puncture and the stuffed puncture initial data violate the constraints. This constraint
   violations leave the black hole horizon. The figure shows the Hamiltonian
   constraint $H$ at time $t=15M$ and $t=30M$. (Left) Puncture initial data, (right) Stuffed puncture initial data.
 }
\end{figure*}

\paragraph*{Single puncture.}
In order to test the damping effect on strong-field evolution, we evolve
puncture initial data~\cite{Brandt:1997tf}.  
While the initial data are constraint- satisfying, a 
constraint-violating wave leaving the hole and propagating outside is
observed in numerical simulations. This feature is generic and not
related to the use of Z4, but observed also in BSSNOK
evolutions, e.g.~\cite{Brown:2008sb}. Note however that the constraint
violation is converging away with resolution, thus not a continuum
feature (the constraint subsystem in Z4c does not have superluminal
speeds). Figure ~\ref{fig:punc:wave} (left panel) shows a
snapshot of the constraint violation leaving the horizon.
During the evolution the biggest violation is instead found at the puncture, 
where the solution is not smooth. 
A priori, the frequency of the initial constraint-violating wave, as
well as the later violation at the puncture, cannot be estimated,  
whereas their amplitude is expected to be ``small'' (in the sense that
it is converging away). From Fig.~\ref{fig:punc:wave} it is 
evident that the frequency of the constraint-violating wave spans a
certain range of frequencies; in terms of the length scale given by the
mass, $M=1$, we mainly observed violation at a peak frequency
$\bar{\nu}\sim0.5$. It can be considered as ``high'' and we expect it to be
damped since it is within the first octave.

\begin{figure*}[t]
 \centering
 \includegraphics[width=.44\textwidth]{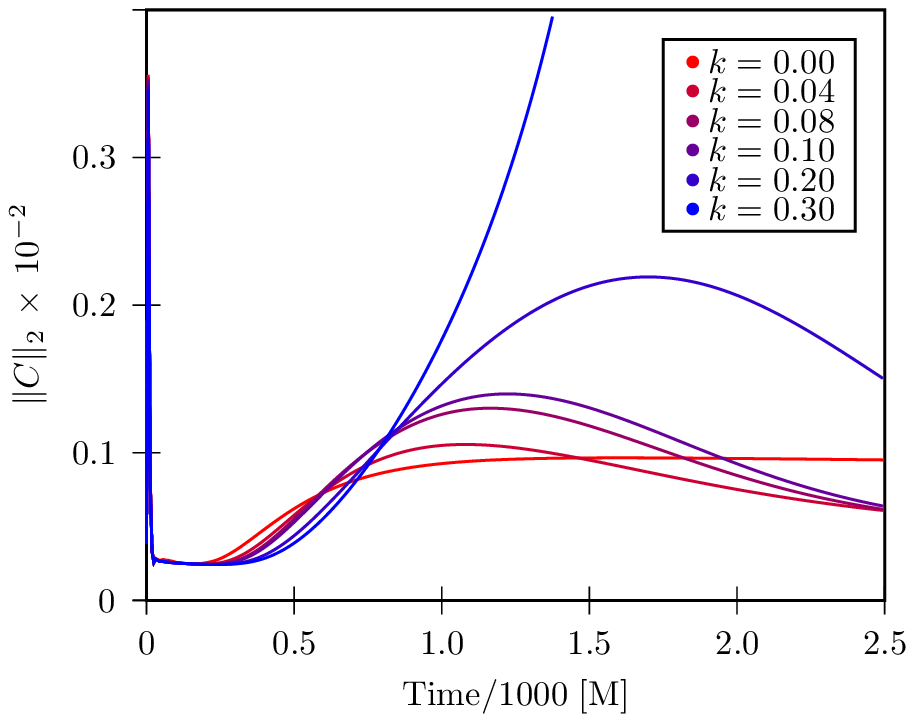}
 \includegraphics[width=.44\textwidth]{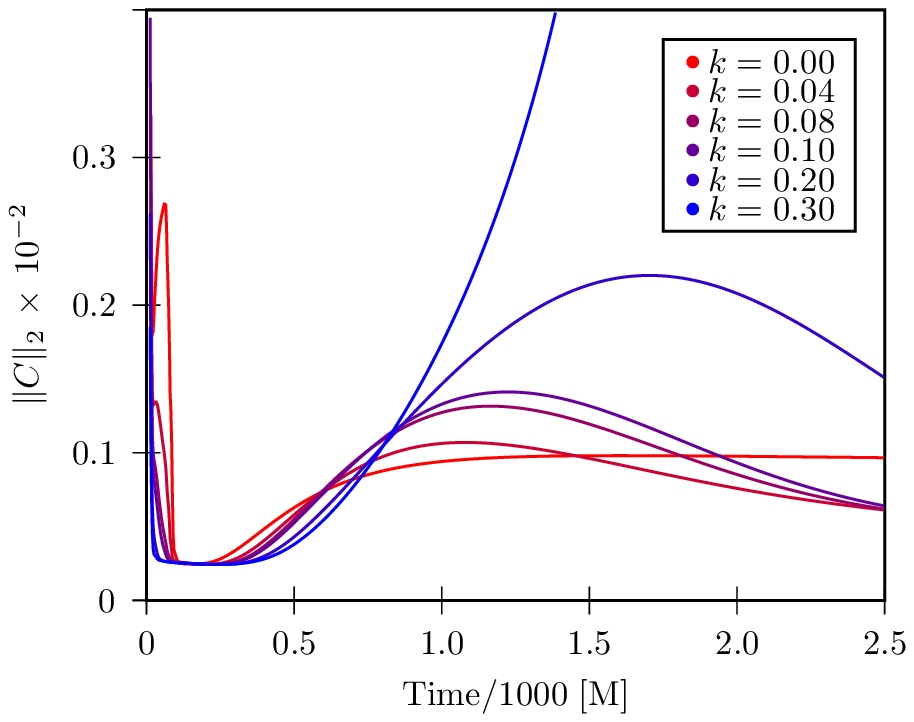}
 \includegraphics[width=.44\textwidth]{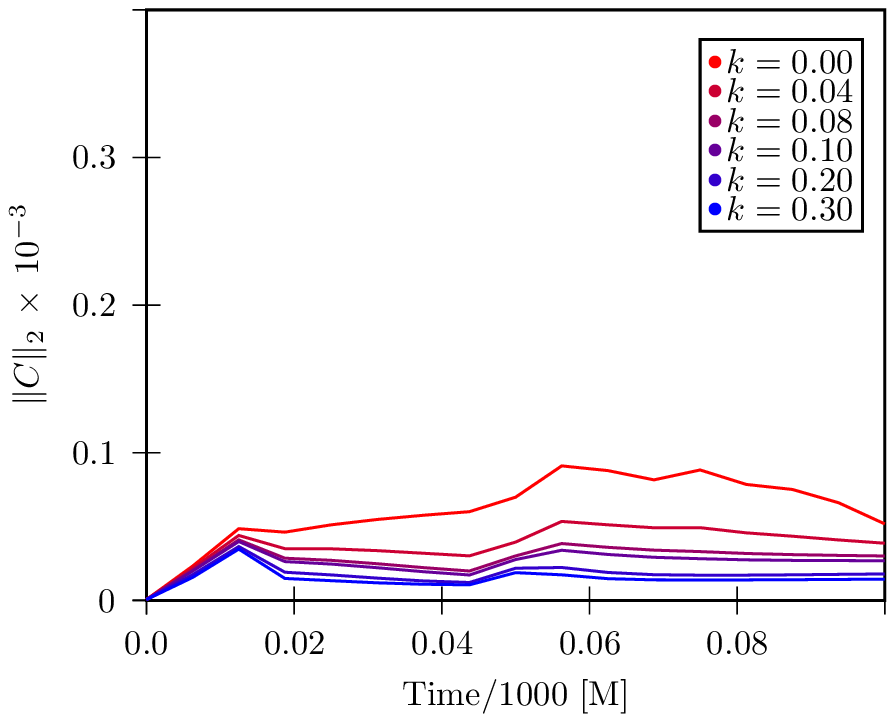}
 \includegraphics[width=.44\textwidth]{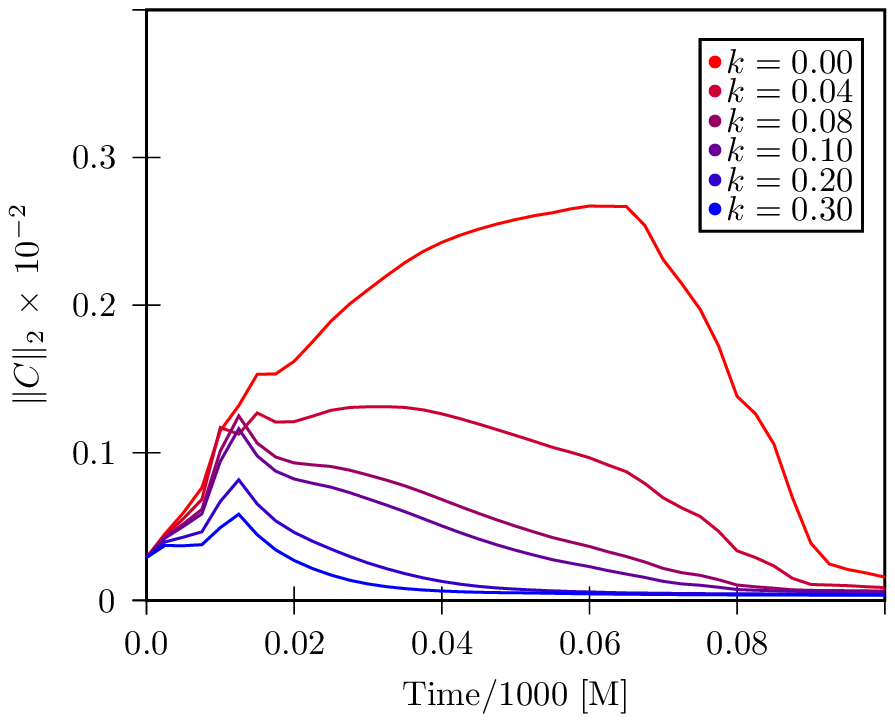}
 \caption{Time evolution of the norm of the constraint monitor.
   For the figures on the left puncture initial data was used, for 
   the figures on the right stuffed puncture initial data. 
   The long time behavior of both cases is the same which can be
   seen in the pictures on the top. The difference between the two
   initial data can be seen in the two figures on the bottom which
   show the norm of the constraint monitor outside the horizon for
   early time. The stuffing introduces a big constraint violation 
   compared to the normal puncture data. This violation is damped away.
   The longtime behavior depends on the value of the damping parameter
   while for damping parameter $k<0.3$ the damping scheme does not 
   cause problems in the evolution, damping parameter $k\ge0.3$ introduce
   dynamics to the system which leads to increasing constraint violation
   and a crash of the code.}
 \label{fig:punc}
\end{figure*}

The numerical evolutions performed with different values of $k$ show
that the use of the damping scheme generically introduces  
a certain dynamics in the constraints, whose values in
space oscillates in time around a small value close to zero.  
The evolution of the L2 norm of the constraint monitor
\begin{align}
 C=\sqrt{H^2+M^i M_i + \Theta^2 + Z^i Z_i}
\end{align}
is reported
in Fig.~\ref{fig:punc} (left panels) for different values of $k$. 
In these tests artificial dissipation is used with $\sigma=0.007$.
In all the cases the norm at early times is dominated by a violation
inside the horizon during the gauge adjustment which leads to the 
trumpet
solution~\cite{Hannam:2006vv,Hannam:2008sg,Garfinkle:2007yt,Brown:2007tb,Brown:2007nt,Thierfelder:2010dv}. The
initial constraint violation wave is  
also propagated out during this phase, and eventually damped depending 
on the value of $k$. The bottom left panel shows the norm outside the 
horizon at early times and highlights the effect of the damping scheme 
for several values of $k$. At later times the amplitude of the 
oscillations in the constraints amplifies around 
$t\sim1.25\times 10^{3}\, M$ as shown in the top left panel. 
Depending on the value of $k$, the amplification is observed to
saturate and damp ($k<0.2$) or to keep on growing, contaminating the
numerical solution ($k>0.2$). In the latter case the code eventually
fails because the boundary conditions implemented~\cite{Ruiz:2010qj}
can not sustain such a large violation.

\begin{figure}[t]
 \centering
 \includegraphics[width=.44\textwidth]{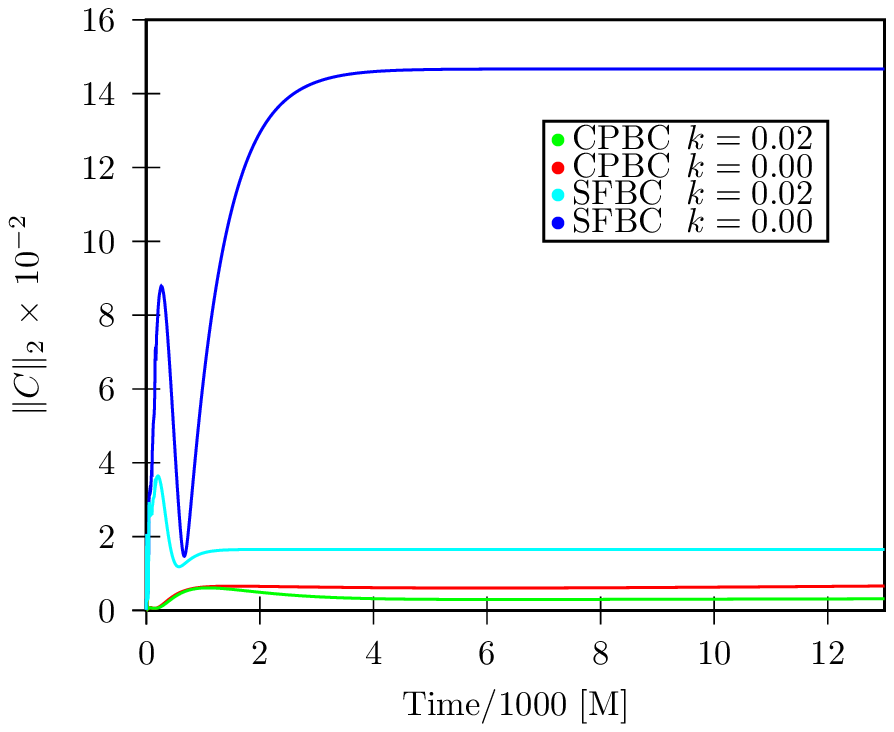}
 \caption{Long-term evolution of a puncture with the Z4c formulation.
          Different boundary conditions were tested with and without
          damping (with $k=0.02$). Sommerfeld boundary conditions lead to
          a high constraint violation, which is effectively suppressed by
          the damping scheme. Using Sommerfeld, the constraint 
          violation is roughly a factor of~$9$ lower with and without damping at 
          late times. Using constraint-preserving boundary conditions, without 
          damping, leads to a late-term constraint violation which is~$20$ 
          times lower than that of undamped Sommerfeld. 
          The damping is less important in the case of constraint-preserving 
          boundary conditions. Using the standard damping value $k=0.02$, the 
          improvement is by only a factor of 2.}
 \label{fig:punc:longterm}
\end{figure}

To assess the relative importance of boundary conditions and the constraint damping 
scheme in our numerical experiments, we performed long evolutions of a 
single puncture with and without both the damping scheme, and either constraint-preserving
~\cite{Ruiz:2010qj} or Sommerfeld boundary conditions. The outer 
boundary was placed at $50M$. The results are presented in 
Fig.~\ref{fig:punc:longterm}. We find that the use of constraint-preserving 
boundary conditions is more important than that of the damping scheme in avoiding 
violations. Although the use of the damping scheme with $k=0.02$ reduces the 
violation by a factor of 9 when Sommerfeld conditions are used. This test is 
not expected to be representative of more general scenarios 
in which the outer boundary is placed farther out with the same resolution, since 
then, experimentally we find that a smaller constraint violation interacting with 
the Sommerfeld condition results in smaller reflections. On the other hand, since 
the constraint-preserving conditions are found to converge numerically, and the 
Sommerfeld conditions do not, it is expected that at some resolution the constraint 
violation induced by the Sommerfeld conditions will become dominant, even if the 
outer boundary is placed far out. This has been recently pointed out for the 
case of 3d matter simulations and of BSSNOK in~\cite{Thierfelder:2011yi}.

\paragraph*{Stuffed puncture.} A second series of tests performed is
the evolution of puncture initial data stuffed in the black-hole
interior~\cite{Faber:2007dv,Brown:2007pg,Brown:2008sb}. 
The hole has been stuffed inside the horizon $r_h=M/2$ at
$r_{\mathrm{ex}}=0.475$ using a fourth-order polynomial in the
conformal factor
\begin{equation}
 \psi(0,r)=2.97368 - 5.83175 \left(\frac{r}{M}\right)^2 + 7.75413
 \left(\frac{r}{M}\right )^4
 \ .
\end{equation}
The polynomial matches the puncture data at $r_{\mathrm{ex}}$ up to
the second derivatives. The initial data are clearly constraint-violating 
and also show the outgoing constraint-violation wave
Fig.~\ref{fig:punc:wave} (right panel).The evolution of the constraint 
monitor for different $k$ is reported in the right panels of
Fig.~\ref{fig:punc}.   
Only quantitative differences with respect to the puncture case are
observed.
As demonstrated in the right bottom panel of Fig.~\ref{fig:punc} (note 
the difference in the scale with respect the left panel),   
the damping scheme is again effective in reducing the outgoing
constraint violation during the initial adjustment. 
At late times the situation is completely analogous to the puncture
evolution and we do not repeat the description, see top right panel.

\begin{figure*}[ht]
 \centering
 \includegraphics[width=.44\textwidth]{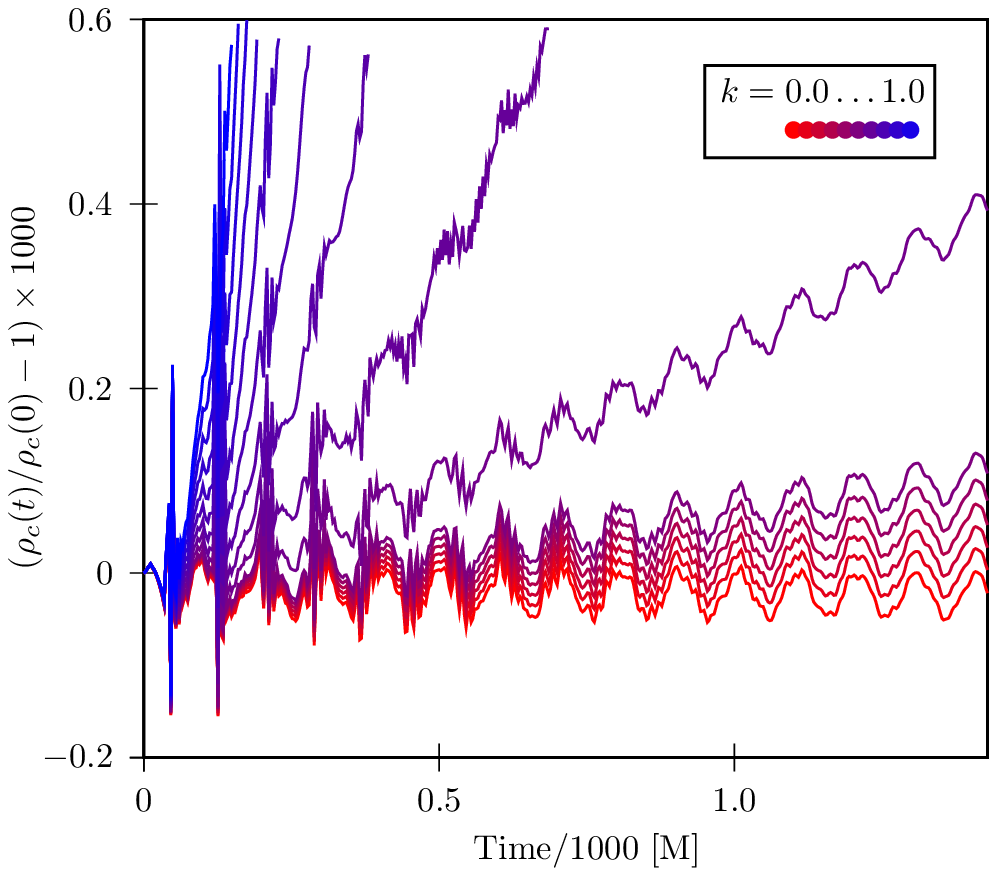}
 \includegraphics[width=.44\textwidth]{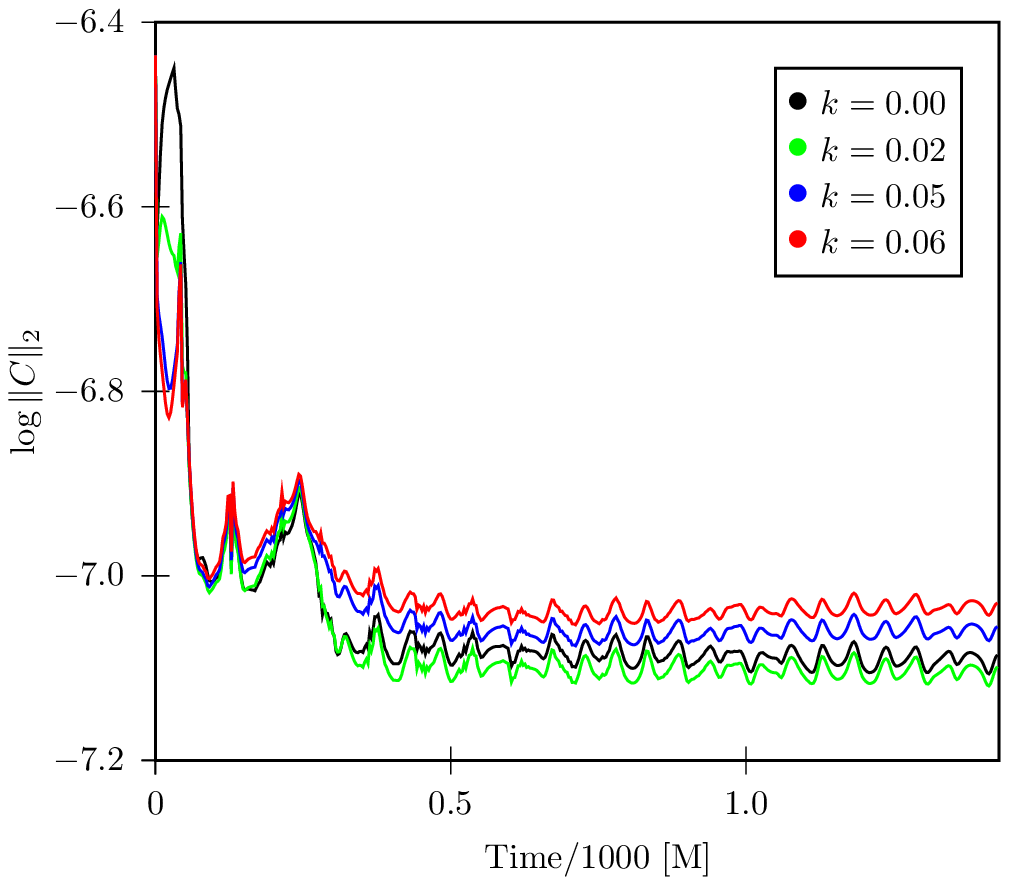}
 \caption{(Left) Time evolution of central rest-mass density for a
   stable star obtained using different values of $k$.
   (Right) Time evolution of the constraint monitor corresponding to a
   subset of the evolutions of the right panel.}
 \label{fig:starcol}
\end{figure*}
 
\paragraph*{Stable star.} As a final test, we present a study of the
influence of the damping scheme on the evolution of a stable
equilibrium model of a compact star of mass $M=1.4\,M_\odot$ described by 
the ideal gas equation of state.
Initial data are the same as those employed in previous
works~\cite{Bernuzzi:2009ex,Ruiz:2010qj} and constraint-satisfying.
Furthermore they provide the exact solution for the evolution
problem since the system is static.
At the typical resolutions employed (and without damping, $k=0$) stable
evolutions are obtained for about $10~ms$~($t\sim1400\,M$).
Truncation errors trigger radial oscillation in the star, which are
small amplitude, low frequency in space constraint-violating modes, 
$\nu\sim1/(2R)$ where $R$ is the star coordinate radius.

In previous investigations on stable stars, which considered only one value of 
the damping parameter, we found that the constraint damping terms have a negligible 
effect on the dynamics, while constraint propagation made a large difference with 
the equivalent BSSNOK simulations~\cite{Bernuzzi:2009ex}. The new results discussed 
here confirm the previous ones for the specific $k$ considered, but they also show 
that for certain values of the damping parameter the constraint damping terms are 
not beneficial in the long-term evolutions. 

Figure~\ref{fig:starcol} shows the evolution of the central rest-mass
density (left) and of the L2 norm of the constraint monitor
(right) for different values of $k$ employed in the damping scheme.
For~$k\ge0.3$ the constraint damping amplifies the radial oscillations
and drives the star to collapse. Large constraint violations 
indicate the departure from the constraint-satisfying solution space 
(for clarity they are not shown in the right panel).
Smaller values of $k\leq0.06$ produce instead an effective
constraint damping at early times (see right panel
Fig.~\ref{fig:starcol}). In the long-term however, 
the evolution without damping scheme ($k=0$) is always
preferable to the evolutions with $k\ge0.03$. In the latter cases a
long-term growth is observed, similar to that seen in the puncture simulations. 
The value $k=0.02$ leads to a constraint damped evolution, but its
effect is almost negligible.
More importantly, this choice is not robust in other tests performed. 
In both simulations employing a different equation of state, specifically a
polytrope which produces different truncation errors at the surface of
the star, and in the migration test of~\cite{Bernuzzi:2009ex} we were
not able to identify a value of $k$ which leads to an efficient
constraint damping.
We presented here in detail only the test with the most varied outcome.

While the static stable star is a delicate 
test (every small perturbation causes a departure from the Einstein 
solution), it provides a specific relevant example in which the constraint 
damping fails for a large choice of damping parameters.
This indicates that the use of the constraint damping scheme without 
specific investigations is potentially dangerous.

\section{Conclusion}
\label{sec:Conclusion}

In order to expand the body of evidence that a conformal 
decomposition of the Z4 formulation of general relativity~\cite{Bona:2003fj} 
may be a useful tool for numerical relativity we have presented 
a detailed study of the effect of the constraint damping scheme
of Gundlach et al.~\cite{Gundlach:2005eh}. 

We have attempted to answer three questions, which we address 
here specifically: 

(i).~{\it Under what conditions can the theoretically 
  predicted damping rates be recovered in the numerical approximation?}
By studying the evolution of parametrized constraint violating-perturbations 
on top of flat space, we first found that the predicted 
damping rates of~\cite{Gundlach:2005eh} are recovered for well-resolved 
high-frequency constraint violations. Varying the frequency of the 
constraint violation, we found that the analytically predicted 
exponential decay is maintained over a large, three-octave, 
range. The cut-off in the effectiveness of the scheme 
occurs over a small range at low frequencies. On grid noise, 
unsurprisingly, we find that the predicted damping rates are not 
recovered, although the combination of damping and artificial 
dissipation does help to suppress constraint violations. The 
intuitive explanation for this is that artificial dissipation
aliases the grid noise to lower frequencies which are 
well-resolved, and on which the damping scheme is effective. Finally
we increased the amplitude of the constraint violation. At amplitudes 
above $A\simeq 0.1$ the damping scheme becomes increasingly less 
effective, after which numerical integration is often not 
possible, either with or without constraint damping.

(ii).~{\it How effective is the damping scheme in astrophysically 
  relevant spacetimes?}
For this part of the investigation we began
by evolving a single puncture black hole. We find that the constraint damping 
scheme suppresses constraint-violating numerical error 
leaving the black hole horizon, but that it generically introduces a 
dynamical behavior to the constraints. The suppression of the 
violation leaving the horizon is furthermore not dramatic. For 
reasonable values of the damping parameter a factor of about 2 or 
3 is gained in the norm of the constraint violation. If the damping 
parameters are chosen too large, the dynamical behavior induced by 
the scheme causes a large constraint violation to hit the outer 
boundary, which in our tests was placed at $50M$, eventually causing 
a code failure, which we think it may be possible to avoid by including 
constraint damping terms in the constraint-preserving boundary 
conditions~\cite{Ruiz:2010qj} appropriately. Since mixed puncture-black-hole 
neutron-star initial data are not readily available, 
``black-hole stuffing'' has been proposed~\cite{Faber:2007dv,Brown:2007pg,Brown:2008sb}. 
Therefore to investigate the likely effect of the constraint damping scheme in 
mixed binary evolutions, we evolved a single puncture with a constraint-violating 
interior. Here we find qualitatively the same behavior as 
in the single puncture evolutions, the only difference being that the 
size of the constraint-violating numerical error leaving the black-hole
is larger. Finally on the question of astrophysically relevant 
spacetimes evolutions of a static star were performed. 
Here we find that using the damping scheme is generically of minor
benefit and can cause an unphysical collapse.

(iii).~{\it In practical applications 
  what are reasonable values for the constraint damping coefficients?}
Our flat-space tests demonstrate that higher values of the damping 
parameters are preferable, because then faster rates of exponential 
damping are achieved. On the other hand, since our evolutions of compact 
objects suffer from severe problems when the damping parameters 
are chosen too large, we suggest that the damping parameter are 
chosen in the range~$k\in [0,0.1]$ for puncture evolutions while for
matter evolution the safest option to use $k=0$ unless specific
damping tests are performed.

In summary, at least for the spherical symmetric systems studied within 
this work, the following statements about the constraint damping scheme can 
be made: Considering vacuum spacetimes, the damping scheme may be, for carefully 
chosen damping parameters, a useful tool for suppressing constraint violations. 
This is certainly true if there are features in the numerical setup which cause 
large constraint violations, for example, Sommerfeld boundary conditions or constraint-
violating initial data. If no such features are present, the damping scheme is 
not essential and can furthermore affect the physics of the system if the damping 
parameters are taken too large. In the evolution of a static compact star our 
numerical evidence indicates that the damping scheme sometimes leads to a slight 
decrease of constraint violation. On the other hand the damping scheme, in 
combination with some numerical setups, causes growth of the constraints; in the 
special cases we have considered the damping scheme is of marginal use.

\bigskip
\begin{acknowledgements}
The authors would like to thank Bernd Br\"ugmann and Milton Ruiz
for helpful discussions. We also thank the authors 
of~\cite{Alic:2011gg} for valuable comments on the manuscript, 
and, in particular, Carlos Palenzuela for his query on our neutron 
star results. This work was supported in part by DFG grant 
SFB/Transregio~7 ``Gravitational Wave Astronomy,'' the DLR grant
LISA Germany and the DFG Research Training Group 1523/1 
``Quantum and Gravitational Fields''.
\end{acknowledgements}


\bibliographystyle{hunsrt}
\bibliography{refs20120220}
\end{document}